\documentstyle[12pt]{ioplppt}          % use this for preprint style

\begin{document}
\jl{1}

\newcommand \be {\begin{equation}}
\newcommand \ee {\end{equation}}

\title{Localisation in 1D random random walks}

\author{ Albert Compte\dag \ddag, Jean-Philippe Bouchaud\dag}

\address{\dag\ Service de Physique de l'\'etat Condens\'e,
 Centre d'\'etudes de Saclay,  Orme des Merisiers, 
91191 Gif-sur-Yvette C\'edex, France}

\address{\ddag\ Departament de F\'{\ii}sica, F\'{\ii}sica
Estad\'{\ii}stica,
Universitat Aut\`onoma de Barcelona, 08193 Bellaterra, Catalonia,
Spain}

\begin{abstract}
Diffusion in a one dimensional random force field leads to
interesting localisation effects, which we study using the
equivalence with a directed walk model with traps. We show that
although the average dispersion of positions
 $\overline{\langle x^2 \rangle - \langle x \rangle ^2}$ diverges for
long times, the probability that two particles occupy the same site
tends to a finite constant in the small bias phase of the model.
Interestingly, the long time properties of this off-equilibrium,
aging phase is similar to the equilibrium phase of the Random Energy
Model.
\end{abstract}

%
%  Comment if journal format required
%
%\pacs{,,}
%\maketitle

\section{Introduction}

The properties of random walks in random environments can be
markedly different from those of homogeneous random walks
\cite{BG}. For example, the typical distance travelled by a diffusing
particle in an unbiased random force field  in one dimension grows
with time as $x \propto \log^2 t$, instead of the usual $\sqrt{t}$
law \cite{Sinai}. This is due to the fact that the potential energy
typically grows as $\sqrt{x}$, leading to very high barriers which
slow down the progression of the particle. More strikingly, Golosov
has shown that the relative distance between two particles in the
same random force field remains {\it finite} even for large times
\cite{Golosov}, whereas it also grows as $\sqrt{t}$ in a homogeneous
medium. This remarkable classical localisation phenomenon is due to
the fact that the `best' potential minimum which can be reached by
the particles after a long time $t$ is so much better than the
`second best' that all the particles have time to gather there,
before eventually moving to a still better location. 

In the presence of a non-zero average bias $F_0 >0$, several regimes
must still be distinguished. For small enough $F_0$, the mean
position of the particles grows as $t^\mu$, where the exponent $\mu <
1$ is proportional to $F_0$ \cite{BG,Kesten,Derrida,Annals}. Beyond a
critical force, the particles move to the right with a non-zero
velocity. However, the dispersion around the mean velocity is still
anomalous until $\mu$ reaches the value $\mu=2$. Beyond $\mu=2$, the
spreading is  `normal', i.e. Gaussian with a width growing as
$\sqrt{t}$. The question we want to address in this paper is whether
the Golosov phenomenon survives in the presence of a non-zero average
force. We will actually show that different `localisation'
criteria lead to different answers: while the average width of a
packet of particles diverges with time, there is a finite probability
(even at long times) that two particles are nearby in space. The
physical picture is that the density of particles is concentrated on
a finite number of sites, but the relative distance between these
peaks grows with time. As we shall also discuss, there is a strong
analogy between this problem and the low temperature phase of
Derrida's Random Energy Model \cite{REM}.

\section{Model and simulation}

Actually, one can map the long time behaviour of the problem onto
that of a much simpler directed walk model
\cite{BG,Annals,Artfrancais,Aslangul}, where each particle hops to
the right on an even-spaced discrete lattice, with a  site-dependent
hopping rate $W_n$ distributed as:  
\be
\rho(W)=\frac{1}{\Gamma(\mu)} W^{\mu-1} e^{-W}
\label{distr}
\ee

\begin{figure}[bhtp]
% GNUPLOT: LaTeX picture
\setlength{\unitlength}{0.240900pt}
\ifx\plotpoint\undefined\newsavebox{\plotpoint}\fi
\sbox{\plotpoint}{\rule[-0.175pt]{0.350pt}{0.350pt}}%
\begin{picture}(1500,900)(0,0)
\sbox{\plotpoint}{\rule[-0.175pt]{0.350pt}{0.350pt}}%
\put(264,158){\rule[-0.175pt]{0.350pt}{151.526pt}}
\put(264,228){\rule[-0.175pt]{4.818pt}{0.350pt}}
\put(242,228){\makebox(0,0)[r]{0.52}}
\put(1416,228){\rule[-0.175pt]{4.818pt}{0.350pt}}
\put(264,368){\rule[-0.175pt]{4.818pt}{0.350pt}}
\put(242,368){\makebox(0,0)[r]{0.54}}
\put(1416,368){\rule[-0.175pt]{4.818pt}{0.350pt}}
\put(264,507){\rule[-0.175pt]{4.818pt}{0.350pt}}
\put(242,507){\makebox(0,0)[r]{0.56}}
\put(1416,507){\rule[-0.175pt]{4.818pt}{0.350pt}}
\put(264,647){\rule[-0.175pt]{4.818pt}{0.350pt}}
\put(242,647){\makebox(0,0)[r]{0.58}}
\put(1416,647){\rule[-0.175pt]{4.818pt}{0.350pt}}
\put(264,787){\rule[-0.175pt]{4.818pt}{0.350pt}}
\put(242,787){\makebox(0,0)[r]{0.60}}
\put(1416,787){\rule[-0.175pt]{4.818pt}{0.350pt}}
\put(264,158){\rule[-0.175pt]{0.350pt}{4.818pt}}
\put(264,113){\makebox(0,0){0}}
\put(264,767){\rule[-0.175pt]{0.350pt}{4.818pt}}
\put(498,158){\rule[-0.175pt]{0.350pt}{4.818pt}}
\put(498,113){\makebox(0,0){20}}
\put(498,767){\rule[-0.175pt]{0.350pt}{4.818pt}}
\put(733,158){\rule[-0.175pt]{0.350pt}{4.818pt}}
\put(733,113){\makebox(0,0){40}}
\put(733,767){\rule[-0.175pt]{0.350pt}{4.818pt}}
\put(967,158){\rule[-0.175pt]{0.350pt}{4.818pt}}
\put(967,113){\makebox(0,0){60}}
\put(967,767){\rule[-0.175pt]{0.350pt}{4.818pt}}
\put(1202,158){\rule[-0.175pt]{0.350pt}{4.818pt}}
\put(1202,113){\makebox(0,0){80}}
\put(1202,767){\rule[-0.175pt]{0.350pt}{4.818pt}}
\put(1436,158){\rule[-0.175pt]{0.350pt}{4.818pt}}
\put(1436,113){\makebox(0,0){100}}
\put(1436,767){\rule[-0.175pt]{0.350pt}{4.818pt}}
\put(264,158){\rule[-0.175pt]{282.335pt}{0.350pt}}
\put(1436,158){\rule[-0.175pt]{0.350pt}{151.526pt}}
\put(264,787){\rule[-0.175pt]{282.335pt}{0.350pt}}
\put(45,472){\makebox(0,0)[l]{\shortstack{$\overline{Y_2}(t)$}}}
\put(850,68){\makebox(0,0){time ($\times 25.000 \overline{W}^{-1}$)}}
\put(264,158){\rule[-0.175pt]{0.350pt}{151.526pt}}
\put(264,505){\circle{12}}
\put(276,640){\circle{12}}
\put(287,708){\circle{12}}
\put(299,728){\circle{12}}
\put(311,719){\circle{12}}
\put(323,698){\circle{12}}
\put(334,675){\circle{12}}
\put(346,647){\circle{12}}
\put(358,624){\circle{12}}
\put(369,597){\circle{12}}
\put(381,578){\circle{12}}
\put(393,553){\circle{12}}
\put(405,537){\circle{12}}
\put(416,525){\circle{12}}
\put(428,518){\circle{12}}
\put(440,518){\circle{12}}
\put(452,514){\circle{12}}
\put(463,515){\circle{12}}
\put(475,519){\circle{12}}
\put(487,525){\circle{12}}
\put(498,531){\circle{12}}
\put(510,532){\circle{12}}
\put(522,529){\circle{12}}
\put(534,532){\circle{12}}
\put(545,533){\circle{12}}
\put(557,534){\circle{12}}
\put(569,535){\circle{12}}
\put(580,532){\circle{12}}
\put(592,531){\circle{12}}
\put(604,527){\circle{12}}
\put(616,526){\circle{12}}
\put(627,526){\circle{12}}
\put(639,523){\circle{12}}
\put(651,522){\circle{12}}
\put(662,523){\circle{12}}
\put(674,523){\circle{12}}
\put(686,521){\circle{12}}
\put(698,518){\circle{12}}
\put(709,517){\circle{12}}
\put(721,516){\circle{12}}
\put(733,517){\circle{12}}
\put(745,513){\circle{12}}
\put(756,509){\circle{12}}
\put(768,505){\circle{12}}
\put(780,502){\circle{12}}
\put(791,500){\circle{12}}
\put(803,497){\circle{12}}
\put(815,490){\circle{12}}
\put(827,489){\circle{12}}
\put(838,488){\circle{12}}
\put(850,485){\circle{12}}
\put(862,480){\circle{12}}
\put(873,477){\circle{12}}
\put(885,474){\circle{12}}
\put(897,469){\circle{12}}
\put(909,468){\circle{12}}
\put(920,466){\circle{12}}
\put(932,465){\circle{12}}
\put(944,460){\circle{12}}
\put(955,457){\circle{12}}
\put(967,454){\circle{12}}
\put(979,451){\circle{12}}
\put(991,449){\circle{12}}
\put(1002,446){\circle{12}}
\put(1014,445){\circle{12}}
\put(1026,443){\circle{12}}
\put(1038,439){\circle{12}}
\put(1049,438){\circle{12}}
\put(1061,436){\circle{12}}
\put(1073,434){\circle{12}}
\put(1084,432){\circle{12}}
\put(1096,429){\circle{12}}
\put(1108,428){\circle{12}}
\put(1120,426){\circle{12}}
\put(1131,425){\circle{12}}
\put(1143,423){\circle{12}}
\put(1155,423){\circle{12}}
\put(1166,418){\circle{12}}
\put(1178,417){\circle{12}}
\put(1190,417){\circle{12}}
\put(1202,418){\circle{12}}
\put(1213,416){\circle{12}}
\put(1225,414){\circle{12}}
\put(1237,415){\circle{12}}
\put(1248,412){\circle{12}}
\put(1260,411){\circle{12}}
\put(1272,411){\circle{12}}
\put(1284,411){\circle{12}}
\put(1295,411){\circle{12}}
\put(1307,411){\circle{12}}
\put(1319,411){\circle{12}}
\put(1331,409){\circle{12}}
\put(1342,408){\circle{12}}
\put(1354,407){\circle{12}}
\put(1366,406){\circle{12}}
\put(1377,405){\circle{12}}
\put(1389,406){\circle{12}}
\put(1401,404){\circle{12}}
\put(1413,405){\circle{12}}
\put(1424,404){\circle{12}}
\end{picture}
\caption{Time evolution of the quantity $\overline{Y_2}(t)$ in the
simulation of
a directed random model with hopping rates distributed according to a
gamma distribution of index $\mu = 0.4$. The simulation was carried
out in a
20.000 point lattice with 1.000 particles and averaged over 250
disorder samples. Up to the last observation times more than 99\% of
the particles remained within the lattice. This curve shows
that $\overline{Y_2}(t)$ tends to a non-zero constant for asymptotic
times.}
\label{y2t}
\end{figure}
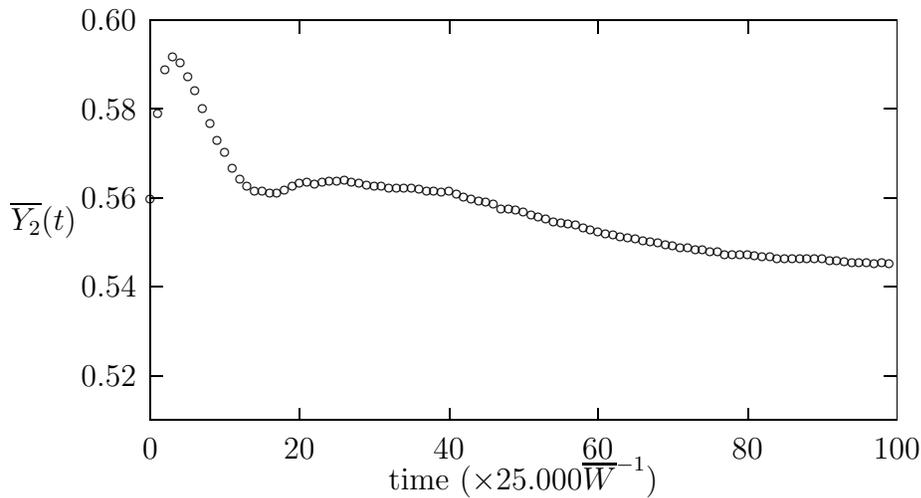

For $\mu < 1$ the average trapping time $\overline{ 1/W }$ is
infinite, this leads to the anomalous behaviour of the particles
average position reported above. The probability to find the particle
on site $n$ obeys the following Master equation:
\be
\frac{dP_n(t)}{dt} = -W_n P_n(t) + W_{n-1} P_{n-1}(t)\label{Master}
\ee 
The properties of equation (\ref{Master}) can be discussed in detail
\cite{Aslangul,LeDoussal}. In particular, one can compute the average
position of the particle, defined as: 
\be
\overline{\langle x(t)\rangle} \equiv \sum_{n=0}^\infty n
\overline{P_n(t)}
\ee
where the overbar denotes the average over the $W$'s. For $\mu <
1$, one easily finds $\overline{\langle x(t)\rangle}= \sin(\pi
\mu)/[\pi \mu \Gamma(\mu+1)] t^\mu$ at large times \cite{BG}. 
One can also compute the average width $\Delta^2$ of the packet,
defined as:
\be
\Delta^2 \equiv \overline{\langle x^2 \rangle - \langle x \rangle^2}
\equiv
\overline{
\sum_{n=0}^\infty n^2 {P_n(t)} - \left(\sum_{n=0}^\infty n
{P_n(t)}\right)^2}
\ee
As shown in \cite{Aslangul}, the width grows to infinity as 
\be
\Delta^2 = C(\mu)  t^{2\mu},
\ee
where $C(\mu)$ is a certain $\mu$ dependent number which can be
explicitely
computed \cite{Aslangul}, and which goes to zero for $\mu=0$.
Naively, this means that the particles' relative positions get
further and further apart (for $\mu > 0$) as $t$ becomes large, at
variance with
Golosov's result for the unbiased case, which shows that $\langle
x^2 \rangle - \langle x \rangle ^2$ remains finite for large $t$.
However, one can still ask the following question: what is the total
probability that two particles,  initially at site $n=0$, occupy the
same site after time $t$ ? This is obtained as 
\be
Y_2(t) = \sum_{n=0}^\infty P_n(t)^2
\ee
The notation $Y_2$ is introduced in analogy with spin-glasses, where
the same question is asked about two copies (replicas) of the same
system in equilibrium, and it measures the probability that these two
copies occupy the same state \cite{MPV}. Note that $Y_2$ is also
often taken to be an  indicator of localisation in quantum problems,
where $P_n =|\psi_n|^2$ is the quantum probability of presence
\cite{quantum}. One can actually study generalized objects, such as:
\be
Y_q(t) = \sum_{n=0}^\infty P_n(t)^q
\ee
which measure the probability that $q$ particles occupy the same
site. We have studied numerically $Y_2$ and $Y_3$ for the problem
defined by equation (\ref{Master}). In figure \ref{y2t}, we show the
average value $\overline{Y_2}(t)$ as a function of $t$ for
$\mu=0.4<1$. 
This quantity clearly tends to a non-zero constant $y_2(\mu)$, which
we plot as a function of $\mu$ in figure \ref{theoexp}, together 
with our theoretical prediction (see below).

\begin{figure}[htbp]
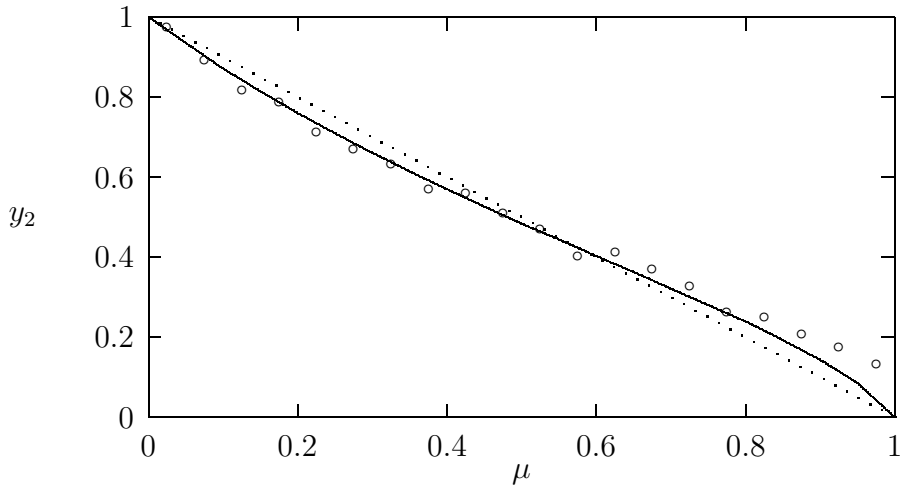

\input fig4
\caption{$y_2$ versus $\mu$ as given by numerical simulations (small circles), the analytic formula obtained
in the text (solid line) and by the equilibrium result of the 
associated Random Energy Model, $y_2=1-\mu$, (dotted line). The agreement
with the theoretical prediction is good, except for values near
$\mu=1$: the observation time being finite,
we expect to overestimate $y_2(\mu)$ more and more the closer one
gets to
$\mu=1$ since the approach to the actual asymptotic value becomes
very slow (logarithmic) for $\mu=1$.}
\label{theoexp}
\end{figure}

\begin{figure}[htbp]
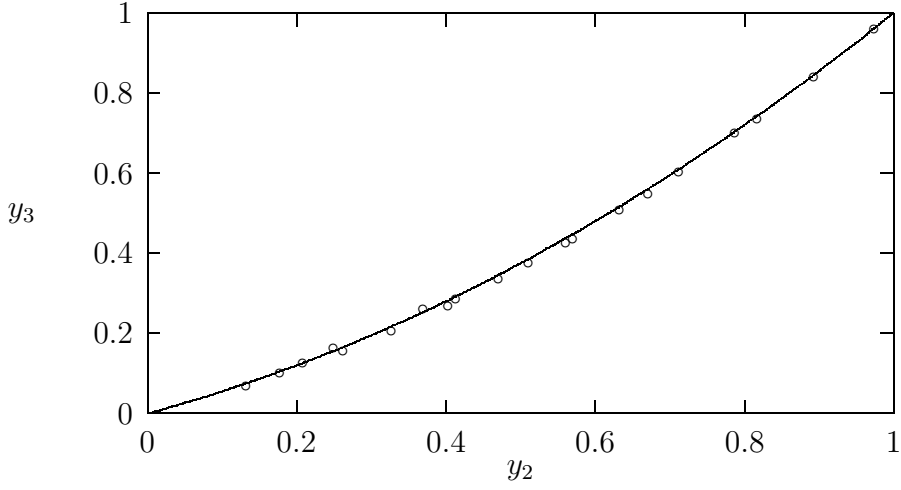

\input fig3
\caption{Representation of $y_3$ versus $y_2$ as obtained from our
simulation (small circles) and as given by the relation $y_3=y_2
(1+y_2)/2$, characteristic of replica symmetry breaking (solid line).
The
agreement is remarkable and hints to some deep relation between the
present 
dynamical problem and the equilibrium phase of disordered systems.}
\label{y2y3}
\end{figure}

We see that $y_2(\mu \to 0)=1$, as expected from
Golosov's results, while we observe the tendency $y_2(\mu \to 1)\to
0$, although  our numerical data is biased in this limit 
%%%%%%%%%%%%% added to suppress figure 2
due to the slow approach to the actual asymptotic value. 
%%%%%%%%%%%%%
In figure \ref{y2y3}, we show the parametric plot of
$y_3(\mu)$ [i.e. the asymptotic value of $\overline{Y_3}(t)$] versus
$y_2(\mu)$.
Interestingly, the resulting curve is seen to be very close to
$y_3=y_2(1+y_2)/2$ obtained within the so-called `one step
replica symmetry breaking' solution of equilibrium random systems.

Before giving a more detailed physical interpretation of these
results, we first turn to an analytic calculation of $y_2(\mu)$.

\section{Analytic derivation of $y_2(\mu)$}

In the trapping model that we study, the time that the particle
sojourns in  the $i$-th trap is given by $t_i=u_i \tau_i$ where
$\tau_i$  is the characteristic trapping time of the $i$-th trap
($\tau_i=W_i^{-1}$) and $u_i$ is an exponentially distributed
variable accounting for the individual thermal behaviour of the
particle. It is then straightforward to write the following equation
for the probability of a particle being at site $n$ at time $t$,
given a realization of the disorder $\{ \tau_i \}$ (or, equivalently,
$\{ W_i \}$):
\be
P_n(t)=\int \prod_i^n du_i \exp\left( -\sum_j^n u_j\right) 
\theta\left(t-\sum_i^{n-1} u_i \tau_i \right) \theta \left(
\sum_i^n u_i\tau_i - t\right)
\label{pdenit}
\ee
$\theta$ being the Heaviside function. Equation (\ref{pdenit}) says
that $P(n,t)$ is the sum of the probabilities of all the possible
thermal histories such that the particle has already done $n-1$ jumps
up to time $t$ but not yet $n$.

We are interested in the probability that two
particles, starting together at $t=0$ remain at the same site
after a sufficiently long time. Therefore, the quantity of interest
is the probability that two particles are at site $n$ at time $t$:
\begin{eqnarray}
\fl P^2_n(t)=\int \prod_i^n du_i dv_i \exp\left(-\sum_j^n (u_j +
v_j)
\right) \theta\left(t-\sum_i^{n-1} u_i \tau_i \right)\theta \left(
\sum_i^n u_i\tau_i - t\right)\nonumber\\
\times 
\theta\left(t-\sum_i^{n-1} v_i \tau_i 
\right) \theta \left(\sum_i^n v_i\tau_i - t\right)
\end{eqnarray}
or, in the Laplace domain,
\be
\fl P^2_n(E)=\int \prod_i^n du_i dv_i \exp\left(-\sum_j^n (u_j + v_j)
\right) \int_{\max\left( \sum_i^{n-1} u_i \tau_i, \sum_i^{n-1} v_i 
\tau_i \right)}^{\min\left( \sum_i^n u_i \tau_i, \sum_i^n v_i 
\tau_i \right)} e^{-Et} dt
\label{laplace}
\ee
Integrating and using $\theta$ functions we can rewrite this 
last expression as
\begin{eqnarray}
\fl E P^2_n(E)= 2\int \prod_i^n du_i dv_i  \left\{ \left[
\exp\left(-E\sum_i^{n-1}
u_i\tau_i\right) -
\exp\left(-E\sum_i^n u_i\tau_i\right) \right]\right.\nonumber\\
\left. \times \theta\left(\sum_i^{n-1} 
(u_i - v_i) \tau_i \right)\theta\left(\sum_i^n 
(v_i - u_i) \tau_i \right) \theta\left(\sum_i^n 
u_i  \tau_i -\sum_i^{n-1} u_i \tau_i \right)\right.\nonumber\\
\left.+\left[ \exp\left(-E\sum_i^{n-1} u_i\tau_i\right) -
\exp\left(-E\sum_i^n v_i\tau_i\right) \right]\theta\left(\sum_i^{n-1} 
(u_i - v_i) \tau_i \right)\right.\nonumber\\
\left. \times \theta\left(\sum_i^n 
(u_i - v_i) \tau_i \right)\theta\left(\sum_i^n 
v_i  \tau_i -\sum_i^{n-1} u_i \tau_i
\right)\right\}\exp\left(-\sum_j^n (u_j + v_j) \right)
\label{thetas}
\end{eqnarray}
where the three $\theta$'s in each summand implement the maximum
condition in the lower limit of the integral, the minimum condition
in the upper limit and the condition that the upper limit is greater
than the lower limit in (\ref{laplace}), respectively. In
(\ref{thetas}) we have also
used the symmetrical integration with respect to $u_i$ and $v_i$ to
simplify somewhat the expression (whence the factor $2$).

It can now be proved that expression (\ref{thetas}) is equivalent to
a much simpler formula, where only one $\theta$ function appears per
summand and this is accomplished by conveniently renaming $u_i
\leftrightarrow v_i$ as integration variables for some terms and by
bearing in mind
that $u_i$, $v_i$, and $\tau_i$ are all positive, whence the
summations are all monotonous increasing functions of the step number
$n$. This procedure leads to
\begin{eqnarray}
\fl E P^2_n(E)=2 \int \prod_i^n d u_i d v_i \exp \left( -\sum_i^n
(u_i+v_i) \right) \left[ \exp\left( -E \sum_i^{n-1} u_i\tau_i \right)
\theta \left( \sum_i^{n-1} (u_i-v_i) \tau_i \right) \right.
\nonumber \\
\left. -\exp\left( -E
\sum_i^n u_i\tau_i \right) \theta \left(\sum_i^n (v_i - u_i) \tau_i
\right) - \exp\left(-E \sum_i^{n-1} u_i\tau_i \right)\right.
\nonumber\\
\left.  \times \theta \left(
\sum_i^{n-1} u_i\tau_i -\sum_i^n v_i\tau_i\right) + \exp\left(-E
\sum_i^n u_i\tau_i\right) \theta\left(\sum_i^{n-1} v_i\tau_i-\sum_i^n
u_i\tau_i\right) \right]
\label{onetheta}
\end{eqnarray}

In order to proceed we now use the following representation of
the $\theta$ function in the complex plane
\be
\theta(x)=-\frac{\i}{2\pi}\int_{-\infty}^\infty \frac{e^{\i\lambda
x}} {\lambda} d\lambda 
\ee
and (\ref{onetheta}) turns into
\begin{eqnarray}
\fl E P^2_n(E)= -\frac{\i}{\pi} \int_{-\infty}^\infty
\frac{d\lambda}{\lambda}
\int \prod_i^n du_i dv_i \exp\left(-\sum_i^n(u_i+v_i)\right) \left[
\exp\left(-\sum_i^{n-1}[Eu_i-\i\lambda (u_i-v_i)]\tau_i\right)
\nonumber \right.\\
\left.- \exp\left(-\sum_i^n[Eu_i-\i\lambda (v_i-u_i)]\tau_i\right)
-\exp\left(-\sum_i^{n-1}[Eu_i-\i\lambda (u_i- v_i)]\tau_i
\right)e^{-\i\lambda v_n \tau_n} 
\right.\nonumber\\
\left.+\exp\left(-\sum_i^{n-1}[Eu_i-\i\lambda( v_i-
u_i)]\tau_i\right) e^{-(E-\i\lambda)u_n\tau_n}
\right]
\end{eqnarray}
This formulation of the equation permits the factorization, within
each summand, of the contributions of each $du_i dv_i$ so as to get
\begin{eqnarray}
\fl E P^2_n(E)= -\frac{\i}{\pi} \int_{-\infty}^\infty
\frac{d\lambda}{\lambda}
\left[\prod_i^{n-1} \int du dv e^{-u-v} e^{-[Eu-\i\lambda (u-
v)]\tau_i}-\prod_i^n \int du dv e^{-u-v} e^{-[Eu-\i\lambda (v-u)]
\tau_i}  \nonumber \right. \\
 \left.- \int dv_n e^{-(1+\i\lambda\tau_n)v_n}
\prod_i^{n-1} \int du dv e^{-u-v} 
e^{-[Eu-\i\lambda (u- v)]\tau_i}\right. \nonumber\\
\left. +
\int du_n e^{-(1+E\tau_n+\i\lambda\tau_n)u_n} \prod_i^{n-1} \int du
dv e^{-u-v} e^{-[Eu-\i\lambda (v- u)]\tau_i} \right]
\label{factor}
\end{eqnarray}

If we now define the following functions
\begin{eqnarray}
F(E,\lambda)= \int du dv e^{-u-v} \overline{e^{-[Eu-\i\lambda
(u-v)]\tau_i}}\\
G(E,\lambda)= \int du e^{-u} \overline{e^{-(Eu+\i\lambda u)\tau_i}}
\end{eqnarray}
where the bar over the exponential stands for the average over the
possible values of $\tau_i$, the disorder average of (\ref{factor})
is readily written as
\be
\fl E \overline{P^2_n}(E)=-\frac{\i}{\pi} \int_{-\infty}^\infty 
\frac{d\lambda}{\lambda} \left\{ \left[
1-G(0,\lambda) \right]
F(E,\lambda)^n-\left[F(E,-\lambda)
-G(E,\lambda) \right] F(E,-\lambda)^n
\right\}
\label{average}
\ee

We now sum  (\ref{average}) for all the values of $n$ in order to
obtain the function $\overline{Y_2}(E)$:
\be
\overline{Y_2}(E)=\frac{-\i}{\pi} \int_{-\infty}^\infty
\frac{d\lambda}{\lambda}
\left[ \frac{1-G(0,\lambda)}{1-F(E,\lambda)}-
\frac{F(E,-\lambda)-G(E,\lambda)}{1-F(E,-\lambda)} \right]
\label{y2}
\ee

Since $\tau_i=W_i^{-1}$ the functions $F(E,\lambda)$ and
$G(E,\lambda)$ can be expressed in terms
of the distribution of hopping rates $\rho(W)$:
\begin{eqnarray}
\fl F(E,\lambda)=\int_0^\infty dW \rho(W) \frac{W}{W+E-i\lambda}
\frac{W}{W+i\lambda}\nonumber\\
\lo= 1 - \frac{\lambda^2}{E-2\i\lambda}\int_0^\infty dW
\frac{\rho(W)}{W+\i\lambda}-\frac{(E-\i\lambda)^2}{E-2\i\lambda}
\int_0^\infty dW \frac{\rho(W)}{W+E-\i\lambda}
\label{fel}\\
\fl G(E,\lambda)= \int_0^\infty dW \rho(W) \frac{W}{W+E+i\lambda}
\nonumber\\
\lo= 1 - (E+\i\lambda)\int_0^\infty dW \frac{\rho(W)}{W+E+\i\lambda}
\label{gel}
\end{eqnarray}

Using these expressions in  (\ref{y2}) we get, after some algebra,
\be
E \overline{Y_2}(E)=\frac 2\pi\int_0^\infty  \mbox{Re} \left[ 
\frac{ f(i E u)-f(E-\i E u)}{u^2 f(\i E u)
+(1-\i u)^2 f(E-\i E u)} (1-\i u) \right] d u
\label{Y2E}
\ee
the function $f(z)$ being given by the integral
\[
f(z)=\int_0^\infty dW \rho(W)\frac 1{W+z}
\]

The result (\ref{Y2E}) is now straightforwardly applied to the
relevant
distribution of hopping rates $\rho(W)$. For instance, it is
reassuring to see that for a non-disordered lattice, $\rho(W) =
\delta (W-W_0)$, equation (\ref{Y2E}) yields:
\[
\overline{Y_2}(t) \simeq \frac 1{2\sqrt{\pi W_0 t}}
\]
which can be obtained directly since the two particles are
independent, and this
means that the probability that the two particles are on the same
site 
is inversely proportional to the typical distance.

For the case that we are exploring here, we focus on a distribution
of the kind (\ref{distr}), whence the function $f(E z)$ turns out to
be
\[
f(E z)=\int_0^\infty dW \frac{W^{\mu-1} e^{-W}}{\Gamma (\mu)}\frac
1{W+E z}\simeq E^{\mu-1} \int_0^\infty dx \frac{x^{\mu-1}}{\Gamma
(\mu)}\frac 1{x+z}
\]
assuming in the last equality that $E$ is sufficiently small so as to
neglect the factor $e^{-E x}$ within the integral. Some further
developments making use of the definition of gamma functions allow
one to get to
\[
f(E z)\simeq \Gamma(1-\mu) E^{\mu-1} z^{\mu -1} \qquad \mbox{as} 
\qquad E \to 0
\]
Introducing this expression in (\ref{Y2E}) we finally obtain
\be
E \overline{Y_2}(E) \simeq  \frac{2}{\pi} \int_0^\infty \mbox{Re}
\left[
\frac{(\i u)^{\mu-1}-(1-\i u)^{\mu -1}}{(1-\i u)^{\mu +1} - (\i
u)^{\mu +1}} (1- \i u) \right] d u
\label{result}
\ee
which is indeed a finite integral when $0<\mu<1$. Equation
(\ref{result}) can now be trivially transformed to the time domain
again and we obtain a $\mu$-dependent constant asymptotic result:
\be
y_2(\mu)=\frac{2}{\pi} \int_0^\infty \mbox{Re} \left[
\frac{(\i u)^{\mu-1}-(1-\i u)^{\mu -1}}{(1-\i u)^{\mu +1} - (\i
u)^{\mu +1}} (1- \i u) \right] d u
\label{result2}
\ee
We have calculated (\ref{result2}) numerically for different
values of $\mu$ in the interval of interest and we have compared the
results to the data obtained from the simulation in figure
\ref{theoexp}. The agreement is quite  good except for values near
$\mu=1$, where the results of the simulation turn less reliable
because of the slow relaxation to the actual asymtotic value in
(\ref{result2}).

\section{Discussion}

\begin{figure}[htbp]
% GNUPLOT: LaTeX picture
\setlength{\unitlength}{0.240900pt}
\ifx\plotpoint\undefined\newsavebox{\plotpoint}\fi
\sbox{\plotpoint}{\rule[-0.175pt]{0.350pt}{0.350pt}}%
\begin{picture}(1500,900)(0,0)
\sbox{\plotpoint}{\rule[-0.175pt]{0.350pt}{0.350pt}}%
\put(264,158){\rule[-0.175pt]{282.335pt}{0.350pt}}
\put(264,158){\rule[-0.175pt]{0.350pt}{151.526pt}}
\put(242,158){\makebox(0,0)[r]{0}}
\put(244,158){\rule[-0.175pt]{4.818pt}{0.350pt}}
\put(242,263){\makebox(0,0)[r]{0.1}}
\put(244,263){\rule[-0.175pt]{4.818pt}{0.350pt}}
\put(242,368){\makebox(0,0)[r]{0.2}}
\put(244,368){\rule[-0.175pt]{4.818pt}{0.350pt}}
\put(242,473){\makebox(0,0)[r]{0.3}}
\put(244,473){\rule[-0.175pt]{4.818pt}{0.350pt}}
\put(242,577){\makebox(0,0)[r]{0.4}}
\put(244,577){\rule[-0.175pt]{4.818pt}{0.350pt}}
\put(242,682){\makebox(0,0)[r]{0.5}}
\put(244,682){\rule[-0.175pt]{4.818pt}{0.350pt}}
\put(242,787){\makebox(0,0)[r]{0.6}}
\put(244,787){\rule[-0.175pt]{4.818pt}{0.350pt}}
\put(264,113){\makebox(0,0){0}}
\put(264,138){\rule[-0.175pt]{0.350pt}{4.818pt}}
\put(557,113){\makebox(0,0){5000}}
\put(557,138){\rule[-0.175pt]{0.350pt}{4.818pt}}
\put(850,113){\makebox(0,0){10000}}
\put(850,138){\rule[-0.175pt]{0.350pt}{4.818pt}}
\put(1143,113){\makebox(0,0){15000}}
\put(1143,138){\rule[-0.175pt]{0.350pt}{4.818pt}}
\put(1436,113){\makebox(0,0){20000}}
\put(1436,138){\rule[-0.175pt]{0.350pt}{4.818pt}}
\put(264,158){\rule[-0.175pt]{282.335pt}{0.350pt}}
\put(1436,158){\rule[-0.175pt]{0.350pt}{151.526pt}}
\put(264,787){\rule[-0.175pt]{282.335pt}{0.350pt}}
\put(45,472){\makebox(0,0)[l]{\shortstack{$P_n$}}}
\put(850,68){\makebox(0,0){$n$}}
\put(264,158){\rule[-0.175pt]{0.350pt}{151.526pt}}
\put(264,158){\usebox{\plotpoint}}
\put(264,158){\rule[-0.175pt]{51.312pt}{0.350pt}}
\put(477,158){\rule[-0.175pt]{0.350pt}{3.132pt}}
\put(477,158){\rule[-0.175pt]{0.350pt}{3.132pt}}
\put(477,158){\rule[-0.175pt]{15.418pt}{0.350pt}}
\put(541,158){\usebox{\plotpoint}}
\put(541,158){\usebox{\plotpoint}}
\put(541,158){\rule[-0.175pt]{4.577pt}{0.350pt}}
\put(560,158){\usebox{\plotpoint}}
\put(560,159){\usebox{\plotpoint}}
\put(561,158){\rule[-0.175pt]{5.782pt}{0.350pt}}
\put(585,158){\rule[-0.175pt]{0.350pt}{0.964pt}}
\put(585,158){\rule[-0.175pt]{0.350pt}{0.964pt}}
\put(585,158){\rule[-0.175pt]{56.852pt}{0.350pt}}
\put(821,158){\rule[-0.175pt]{0.350pt}{132.013pt}}
\put(821,158){\rule[-0.175pt]{0.350pt}{132.013pt}}
\put(821,158){\rule[-0.175pt]{3.373pt}{0.350pt}}
\put(835,158){\usebox{\plotpoint}}
\put(835,158){\usebox{\plotpoint}}
\put(835,158){\rule[-0.175pt]{23.126pt}{0.350pt}}
\put(931,158){\rule[-0.175pt]{0.350pt}{54.684pt}}
\put(931,158){\rule[-0.175pt]{0.350pt}{54.684pt}}
\put(931,158){\rule[-0.175pt]{1.204pt}{0.350pt}}
\put(936,158){\rule[-0.175pt]{0.350pt}{0.964pt}}
\put(936,158){\rule[-0.175pt]{0.350pt}{0.964pt}}
\put(936,158){\rule[-0.175pt]{35.653pt}{0.350pt}}
\put(1084,158){\rule[-0.175pt]{0.350pt}{1.927pt}}
\put(1084,158){\rule[-0.175pt]{0.350pt}{1.927pt}}
\put(1084,158){\rule[-0.175pt]{7.227pt}{0.350pt}}
\put(1114,158){\usebox{\plotpoint}}
\put(1114,158){\usebox{\plotpoint}}
\put(1114,158){\rule[-0.175pt]{18.790pt}{0.350pt}}
\put(1192,158){\usebox{\plotpoint}}
\put(1192,158){\usebox{\plotpoint}}
\put(1192,158){\rule[-0.175pt]{9.636pt}{0.350pt}}
\put(1232,158){\rule[-0.175pt]{0.350pt}{27.703pt}}
\put(1232,158){\rule[-0.175pt]{0.350pt}{27.703pt}}
\put(1232,158){\rule[-0.175pt]{1.686pt}{0.350pt}}
\put(1239,158){\usebox{\plotpoint}}
\put(1239,158){\rule[-0.175pt]{16.381pt}{0.350pt}}
\put(1307,158){\usebox{\plotpoint}}
\put(1307,158){\usebox{\plotpoint}}
\put(1307,158){\rule[-0.175pt]{4.818pt}{0.350pt}}
\put(1327,158){\rule[-0.175pt]{0.350pt}{1.927pt}}
\put(1327,158){\rule[-0.175pt]{0.350pt}{1.927pt}}
\put(1327,158){\rule[-0.175pt]{11.081pt}{0.350pt}}
\put(1373,158){\rule[-0.175pt]{0.350pt}{26.981pt}}
\put(1373,158){\rule[-0.175pt]{0.350pt}{26.981pt}}
\put(1373,158){\rule[-0.175pt]{15.177pt}{0.350pt}}
\end{picture}
\caption{Distribution of probability after a time $t=7 \times 10^{10}
\overline{W}^{-1}$ for a particular sample of disorder in our 1D
directed random walk model with $\mu = 0.4$. The simulation was done
with $1.000$ particles in a lattice of $20.000$ sites.}
\label{profile}
\end{figure}
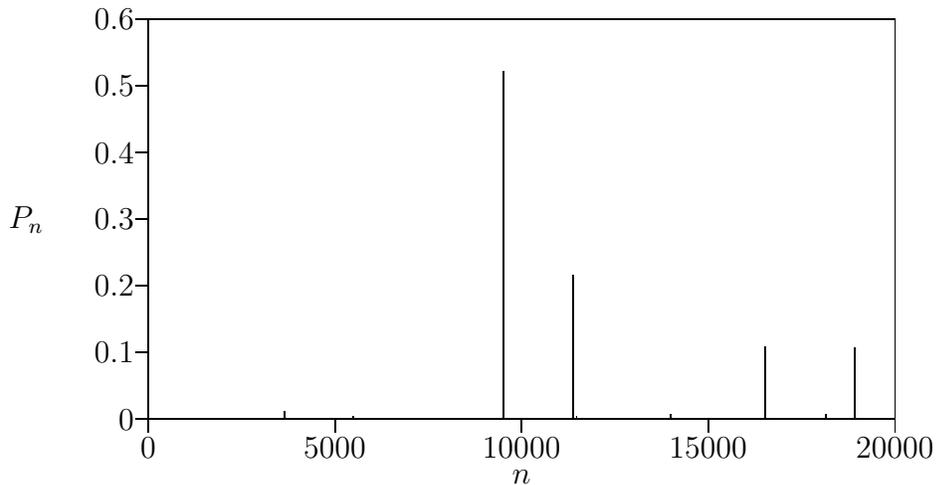

How can one reconcile the fact that, at the same time, the typical
distance between two particles grows with time (as $t^\mu$) and that
the probability to find them at the same site tends to a finite
constant? The physical picture is that of figure \ref{profile},
where the probability $P_n(t)$ is shown for a single sample and a
fixed $t$. 
One sees that this probability distribution is made of several sharp
peaks that gather a finite fraction of the particles. However, the
position of these peaks is scattered on a region of space of width
$t^\mu$. As time progresses, the position and relative weights
of these peaks of course change, but at any given (large) time only a
finite number of peaks, corresponding to very large trapping times,
contain most of the particles. This is clearly related to the fact
that, for a given particle, most of its life was spent in the deepest
`trap' encountered up to time $t$ \cite{WEB,Bardou}. This behaviour
is typical of the L\'evy sums: when $\mu < 1$ the sum of individual
trapping times is dominated by the largest one. There is also a
strong connection with the physics of the Random Energy Model in its
low temperature (glassy) phase. The distribution of Boltzmann weights
(which are also the residence times within each state) is there again
a power-law with an exponent $\mu < 1$, whence only a
finite (but random) number of states contribute to the full
partition function \cite{Derrida2,BM}, and the probability that two
independent copies of the same system occupy the same state is finite
(and equal, on average, to  $y_2=1-\mu$). As discussed in detail in
\cite{BM}, this is in turn related to `replica symmetry breaking'.
All the $y_q$'s can be computed and one finds, in particular,
$y_3=y_2(y_2+1)/2$. As explained above, we find analytically that
$y_2 \neq 1-\mu$ in the dynamical model, which means that the system
can never be considered in equilibrium, although the dynamics gets
slower and slower with time \cite{Review}. At the same time, however,
the equilibrium relation 
$y_3=y_2(y_2+1)/2$ appears to be fulfilled (see figure 3), suggesting
that
some kind of pseudo-equilibrium can be defined, for which equilibrium 
methods such as the replica method could be applied. It would be
interesting
to extend the method of the previous paragraph to calculate all the
$y_q$'s exactly, and to check whether they agree with the replica
prediction.

Finally, it is interesting to note that the above biased model
exhibits aging effects when $\mu < 1$
\cite{Vinokur,WEB,Monthus,Laloux}. In this context, a classification
of different aging models was proposed in \cite{Barrat}, in terms of
the asymptotic `clone overlap' function. The idea is to look at the
evolution of two identical systems (replicas), driven by the same
thermal noise until $t=t_w$, and by independent thermal noise for
later times. The two replicas can either separate with time (type I
aging), or remain close even after infinite time (type II aging). One
sees from the above example that, depending on the way in which one
measures the `closeness' of the two particles, one concludes
differently. This situation is reminiscent of the quantum
localisation model introduced in \cite{Cizeau}, where states are (in
certain regions of parameter space) both extended and localised,
depending on the property which is studied. The physical nature of
these quantum mixed states is actually very similar to the one
discussed
above. 

\ack
We thank A. Baldassarri and M. M\'ezard for useful discussions.
AC wants to express his gratitude to the SPEC for their warm
hospitality at CEA-Saclay and to the Direcci\'o General de Recerca of
the Generalitat de Catalunya for financial support within the
programme FPI. Further funding  has been obtained from the DGICyT of
the Spanish Ministry of Education under grant No PB94-0718.

\end{document}